\documentclass[twocolumn, amsmath,superscriptaddress, nohypertex, amsfonts,prb]{revtex4}
\usepackage{graphicx}
\usepackage{url}
\pdfoutput=1
\begin{document}

\title{The Poole-Frenkel laws and a pathway to multi-valued memory}

\author{Maria Patmiou}
\email{maria.patmiou@rockets.utoledo.edu}
\affiliation{Department of Physics and Astronomy, University of Toledo, Toledo, OH 43606, USA}
\author{D. Niraula}
\email{dipesh.niraula@rockets.utoledo.edu}
\affiliation{Department of Physics and Astronomy, University of Toledo, Toledo, OH 43606, USA}
\author{V. G. Karpov}
\email{victor.karpov@utoledo.edu}
\affiliation{Department of Physics and Astronomy, University of Toledo, Toledo, OH 43606, USA}

\date{\today}

\begin{abstract}

We revisit the mechanism of Poole-Frenkel non-ohmic conduction in materials of non-volatile memory. Percolation theory is shown to explain both the Poole and Frenkel dependencies corresponding to the cases of respectively small and large samples compared to the correlation radii of their percolation clusters. The applied bias modifies a limited number of microscopic resistances forming the percolation pathways. That understanding opens a pathway to multi-valued non-volatile memory and related neural network applications.

\end{abstract}
\maketitle
\section{Introduction}\label{sec:intro}
The non-ohmic conductivity in various materials is often described  in terms of the Poole \cite{poole1916,hill1971} or Frenkel \cite{frenkel1938} laws, (or under the unifying names Poole-Frenkel (PF) or Frenkel-Poole), in the forms,
\begin{equation}\label{eq:P}J\propto \exp(C_P{\cal E}/(kT)) \quad {\rm Poole\ law}\end{equation}
\begin{equation}\label{eq:F}J\propto\exp(C_F\sqrt{{\cal E}}/(kT))\quad {\rm Frenkel \  law}\end{equation}
where $J$ is the current, ${\cal E}$ is the electric field, $C_P$ and $C_F$ are two parameters defined below, $k$ is the Boltzmann's constant, and $T$ is the temperature. The total number of PF related observations exceeds 1300 for the past decade. Recently, phase change memory (PCM) and resistive random access memory (RRAM) structures have been PF objects of significant practical interest. Here, we limit ourselves to their related data.

\begin{figure}[b!]
\includegraphics[width=0.52\textwidth]{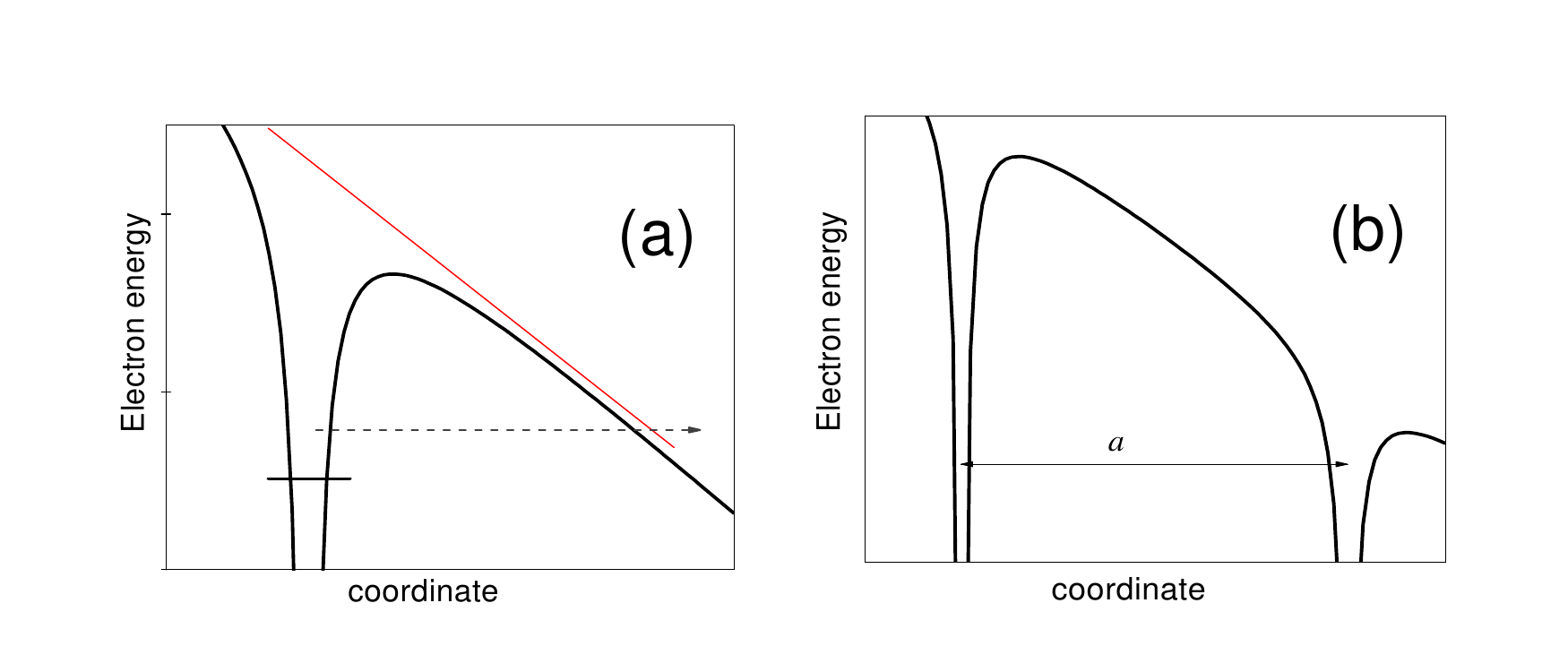}
\caption{(a) Field induced decrease in ionization energy of a coulomb center; horizontal arrow represents possible tunneling ionization. (b) Field induced decrease in transport barrier between two coulomb centers attributable to Poole law. \label{Fig:PF}}
\end{figure}

Eq. (\ref{eq:F}) was originally related to the
decrease in the ionization energy of a single coulombic center in the direction of an applied field \cite{frenkel1938} as illustrated in Fig. \ref{Fig:PF} (a). It predicts $C_F=2\sqrt{q^3/\varepsilon}$ where $q$ is the elemental charge and $\varepsilon$ is the dielectric permittivity. A similar interpretation of PF law was attempted for hopping conduction. \cite{aladashvili1989} A rigorous analysis \cite{perel} shows that Frenkel theory is limited to low $T$ (compared to the characteristic Debye temperature) and weak enough fields ${\cal E}$ not allowing tunneling ionization.

The existing explanation of Poole law \cite{hill1971,ielmini2007,nardone2012} is based on the model of a lattice of equidistant coulomb centers whose overlapping potentials set transport barriers  illustrated in Fig. \ref{Fig:PF} (b), yielding $C_P=aq/2$ where $a$ is the inter-center distance. It is vulnerable to the effects of random fluctuations in distances $a$ causing variations of barrier heights and exponentially broad distribution of transition rates. Percolation theory forms a proper framework for analyzing these types of systems.

Here, we propose a theory that explains both (\ref{eq:P}) and (\ref{eq:F}) in terms of percolation conduction depending on sample size. Its practical application is that bias-induced modifications of the microscopic resistances in a percolation cluster are non-volatile with PCM and RRAM  materials, which opens a pathway to multi-valued memory.

\section{Percolation analysis}\label{sec:perc}

\begin{figure}[t!]
\includegraphics[width=0.5\textwidth]{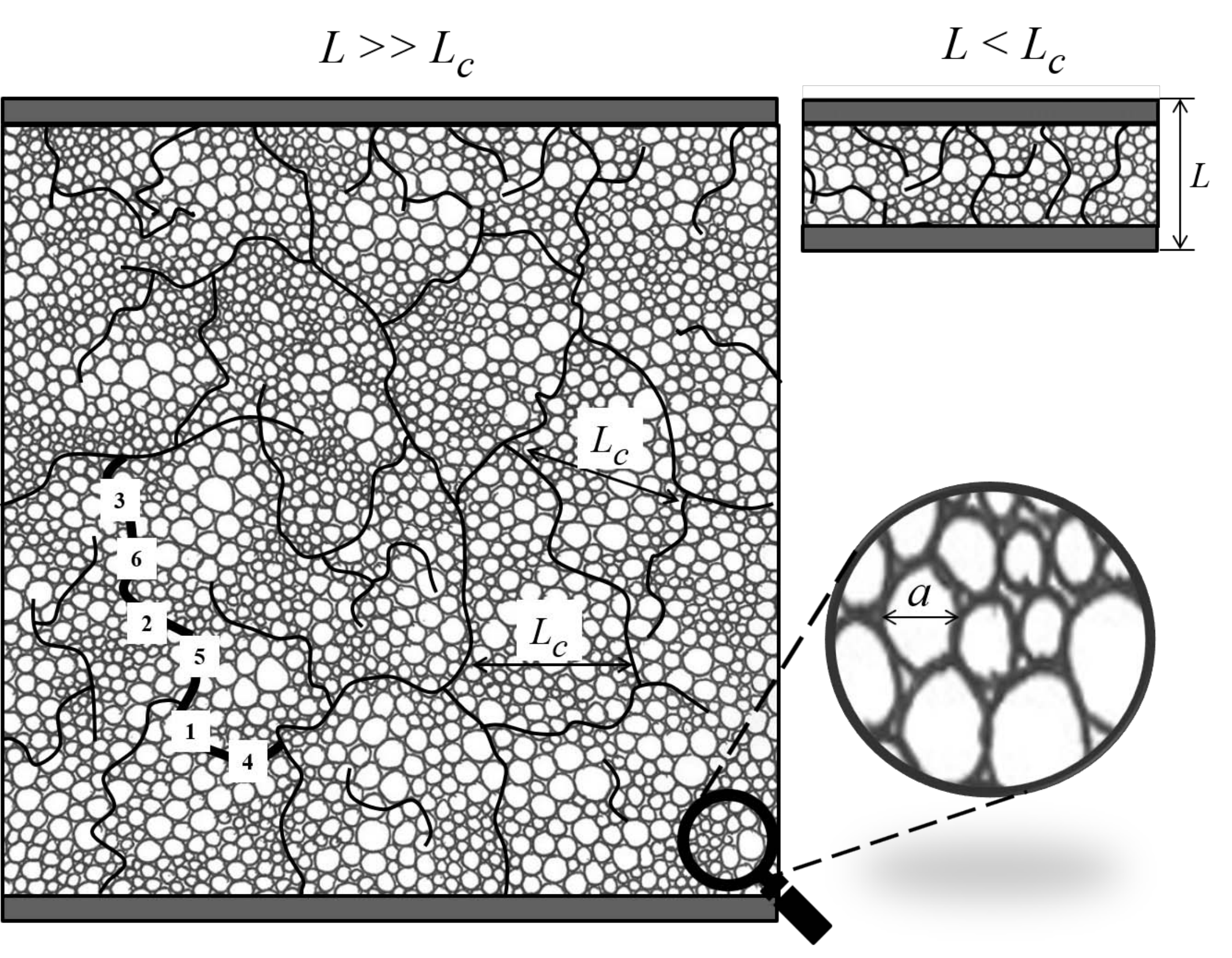}
\caption{A sketch of conductive pathways in large (left) and small (right) samples of a disordered material qualitatively representative of polycrystalline semiconductors or granular metals. Numbers 1-6 in represent random resistors in descending order. \label{Fig:PF_perc}}
\end{figure}

We recall \cite{efros} that the concept of percolation conduction is relevant for disordered systems formed by multiple exponentially different resistors, such as inter-center resistances in hopping transport, barrier resistances in polycrystalline semiconductors, or granular metals. For such systems the conductivity is dominated by the subsystem of minimally strong resistors sufficient to form an everywhere connected cluster (`infinite cluster'). Qualitatively, the electronic conduction in such a cluster is similar to the percolation of water through a system of globally connected channels in a mountain terrain. The cluster is characterized by the correlation radius $L_c$ determining its average mesh size as illustrated in Fig. \ref{Fig:PF_perc}. A system of large linear dimensions $L\gg L_c$ is effectively uniform.

Each bond of the percolation cluster consists of a large number ($i=1,2,...,N\gg 1$) of exponentially different random resistors, all exhibiting non-ohmicity due to the field induced suppression of their corresponding activation barriers $V_i$,
\begin{equation}\label{eq:nonohm}J=J_0\exp\left(-\frac{V_i}{kT}\right)\sinh\left(\frac{eU_i}{2kT}\right)\end{equation}
where $U_i={\cal E}_ia$ is the voltage applied to the barrier, $a$ and ${\cal E}_i$ are respectively the barrier width and local electric field. The random quantities $\xi _i\equiv V_i/kT$ are uniformly distributed in the interval $(0,\xi _m )$ where the maximum barrier $V_m=kT\xi _m$ is determined by the requirement that resistors with $V\leq V_m$ form the infinite cluster.

In a significantly non-ohmic regime, the current can be represented as
\begin{equation}\label{eq:nonohm1}J=J_0\exp[-\xi _m+f({\cal E})]\end{equation}
where ${\cal E}=U/l$ is the macroscopic electric field corresponding to voltage drop $U$ across a sample of length $l$. The theories  of non-ohmic percolation conduction \cite{shklovskii1979,levin1984} determine the shape of $f({\cal E})$. In the case of strong non-ohmicity, the following inequalities must apply,
\begin{equation}\label{eq:ineq}1\ll f\ll \xi _m.\end{equation}
Here the lower limit reflects the condition of strong non-ohmicity, and the upper one guarantees that the conduction barriers are not totally suppressed.

Shklovskii \cite{shklovskii1979} analyzed the non-ohmic conduction in a percolation cluster formed by a random potential in a non-crystalline material as illustrated in Fig. \ref{Fig:PF_perc}. His analysis led to Eq. (\ref{eq:F}) with $C_S=\sqrt{cqaV_0}$ instead of $C_F$, where $a$ and $V_0$ are the characteristic linear and energy scales of the random potential, and $c\sim 1$.

An important conceptual point \cite{shklovskii1979} is that the applied voltage concentrates on the strongest resistor of a percolation bond (resistor 1 in Fig. \ref{Fig:PF_perc}) suppressing it to the level of the next strongest resistor (resistor 2 in Fig. \ref{Fig:PF_perc}), and that the two equally dominate the entire bond voltage drop. Along the same lines, it then suppresses the next-next strongest resistors (3,4,5,.. in Fig. \ref{Fig:PF_perc}), etc.; hence, the percolation cluster changes its structure under electric bias. Specifically,
\begin{equation}\label{eq:corrad}L_c=a\sqrt{\frac{V_0}{q{\cal E}a}}\quad {\rm and} \quad \Delta V_{\cal E}=\sqrt{V_mqa{\cal E}}\end{equation}
where $L_c$ and $\Delta V$ are respectively the field dependent correlation radius and maximum barrier decrease in the percolation cluster.

Levin \cite{levin1984} revisited the issue of non-ohmic percolation conduction, negating the conclusion of percolation clusters changing under electric bias.
For the strong field regime, he evaluated
\begin{equation}\label{eq:Levin9}
\overline{u}=\langle u_i\rangle=\frac{(\ln 2U_0)^2}{2\xi _m}\quad {\rm when} \quad 1\ll U_0\ll \exp(\xi _m)\end{equation}
where angular brackets stand for averaging, and
\begin{eqnarray}\label{eq:notations}u_i&=&\frac{eU_i}{2kT}=\sinh ^{-1}\left[\frac{J\exp(\xi _i )}{J_0}\right],\quad \xi _i=\frac{V_i}{kT},\nonumber \\ \xi _m &\equiv\ &\max \{\xi _i\},\quad U_0\equiv u _{\max}= \frac{J\exp(\xi _m )}{J_0}.\nonumber\end{eqnarray}
The averaging is performed with the uniform probabilistic distribution $\rho (\xi _i) =1/\xi _m$, and in the strong field approximation, $\sinh ^{-1}u\approx\ln(2u)$.
Replacing voltage drops across each of the $L_c/a$ resistors with $\overline{u}$ leads at the conclusion that the percolation cluster structure remains intact under all practical voltages and the dependence of Eq. (\ref{eq:F}) does not hold.

\section{Revisiting the percolation analysis}\label{sec:rev}
Here, we note that the average characteristic $\overline{u}$ is insufficient because it may be smaller than its corresponding fluctuations. Indeed, one can evaluate the dispersion, $\overline{(\delta u)^2}=\overline{u^2}-(\overline{u})^2$ with
\begin{eqnarray}\label{eq:avsq}\overline{u^2}=\int _0^{\xi _m}\left\{\sinh ^{-1}\left[\frac{J\exp(\xi )}{J_0}\right]\right\}^2\frac{d\xi}{\xi _m}\approx \frac{(\ln 2U_0)^3}{3\xi _m}.\end{eqnarray}
As a result, the relative dispersion becomes,
$$\Delta u\equiv\frac{\overline{(\delta u)^2}}{(\overline{u})^2}=\frac{4}{3}\frac{\xi _m}{\ln 2U_0}-1=\frac{4\xi_m}{3(\xi_m+\ln(2J/J_0))}-1.$$
Substituting here $J_0/J$ from Eq. (\ref{eq:nonohm1}) yields,
\begin{equation}\label{eq:N}\Delta u=\frac{4 \xi _m}{3(\ln(2)+f)}-1\approx \frac{4 \xi _m}{3f}-1\approx\frac{4\xi _m}{3f}\gg 1\end{equation}
where the latter approximation and inequality follow from Eq. (\ref{eq:ineq}).

We conclude that the characteristic fluctuations in resistor voltages far exceed their average value and thus the consideration \cite{levin1984} based on Eq. (\ref{eq:Levin9}) lacks validity.

It follows that, starting from a certain $N\gg 1$, a bond of more than $N$ resistors will have its average (increasing proportionally to $N$) overweighing the fluctuations (increasing as $\sqrt{N}$). Therefore we can define the sufficient minimum bond length ${\cal L}=aN=4a\xi _m/3f.$
Its corresponding correlation length becomes \cite{efros}
$$L_c=aN^{\nu}=a(4\xi _m/3f)^{\nu}$$
where $\nu\approx 0.9$ is the critical index. In this letter, we neglect for simplicity the difference between $\nu$ and 1, following the earlier approximations.\cite{shklovskii1979}

The electric field is uniform on the scales exceeding $L_c$ and strongly fluctuates across smaller scales. Therefore we accept that the electric field equals its macroscopic value ${\cal E}$ as measured across length $L_c$. Because of the strong variations between the resistance voltages, most of the voltage will drop across just one resistor making the effective field through it, ${\cal E}_{\rm eff}$, stronger than the average by the factor of $N\gg 1$,
\begin{equation}\label{eq:efffield}
{\cal E}_{\rm eff}={\cal E}(4\xi _m/3f).\end{equation}

The factor $f$ can now be defined as
\begin{equation}\label{eq:fdef}
f={\cal E}_{\rm eff}ae/(2kT).\end{equation}
Combining the latter with Eq. (\ref{eq:efffield}) yields the equation for $f$, from which one finds,
\begin{equation}\label{eq:final}
f=\left(\frac{{2\cal E}ae}{3kT}\right)^{\frac{1}{2}}\left(\frac{V_m}{kT}\right)^{\frac{1}{2}}.
\end{equation}
Eq. (\ref{eq:nonohm1}) with $f$ from Eq. (\ref{eq:final}) is tantamount to Eq. (\ref{eq:F}) with the coefficient $C=\sqrt{2qaV_m/3}$ instead of $C_F$. Therefore, our analysis reinstates the original Shklovskii result \cite{shklovskii1979} ($V_0$ and $V_m$ coincide to the accuracy of an insignificant numerical factor \cite{shik}).

\section{Small samples}\label{sec:small}
Consider the case of small samples with thicknesses $L$ below $L_c$, i. e. beyond the domain of percolation theory as depicted in Fig. \ref{Fig:PF_perc} (right). Each conductive chain  has $N=L/a\gg 1$ random resistors with total voltage drop $\sum U_i=U$. Using the notations of Sec. \ref{sec:perc}, one gets $u_i=\xi _i+ \ln(2J/J_0)$, and
\begin{equation}\label{eq:xi} \sum_{i=1}^{N}\xi_i + (L/a)\ln(2J/J_0)= qU/2kT.\end{equation}
Here a sum of large number $N$ of random variables $S\equiv \sum _i\xi _i$ is a random quantity obeying the central limit theorem. Hence, the Gaussian distribution,
\begin{equation}\label{eq:f(s)}f(S)=\frac{1}{\sqrt{2\pi\langle (\delta S^2)\rangle}}\exp\big\{-\frac{(S-\langle S\rangle)^2}{2\langle (\delta S^2)\rangle}\big\}\end{equation}
with the average and dispersion given by
\begin{equation}\label{eq:avdis}
\langle S\rangle=\frac{L}{2a}\xi _m, \quad \langle(\delta S)^2\rangle = \frac{L}{12a}(\xi _m)^2\end{equation}
where we have taken into account that $\xi _m\gg 1$.

Expressing $S$ in terms of $J$, Eq. (\ref{eq:f(s)}) yields the log-normal distribution for currents, $\rho (J)= f[S(J)]|dS/dJ|$,
\begin{equation}\label{eq:rhoJ}
\rho (J)= \frac{1}{\sqrt{2\pi}\sigma J}\exp\left[-\frac{(\ln J/J_U)^2}{2\sigma ^2}\right]\end{equation}
with
\begin{equation}\label{eq:avdis1}
\sigma =\sqrt{\frac{a}{12L}}\xi _m,\quad J_U=\frac{J_0}{2}\exp\left(\frac{aqU}{2LkT}-\frac{\xi _m}{2}\right) . \end{equation}

We observe that the maximum distribution corresponds to the current $J_m=J_u\exp(-\sigma ^2)$, which reproduces the Poole law of Eq. (\ref{eq:P}) including its coefficient \cite{hill1971,ielmini2007,nardone2012} $C_P=aq/2$.  The latter coincidence takes place in spite of the fact that $a$ is not the next neighbor distance in a lattice of Coulomb centers as suggested earlier \cite {hill1971,ielmini2007,nardone2012}, but rather the characteristic linear scale of the random potential. Furthermore, it is straightforward to see that the average (rather than most likely) current $\langle J\rangle =\int J\rho (J)dJ$  also follows the Poole law with the same coefficient $C_P$.

Note that the estimate in Eq. (\ref{eq:N}) still remains valid, i. e. the dispersion in the affected resistances exceeds its average. Therefore, the scenario of  sequential elimination of the top down resistors applies here as well.

Several additional comments are in order. First, the conducting channels of Fig. \ref{Fig:PF_perc}  (right) do not have to be rectilinear. Secondly, our coefficient $C_P$ is consistent with the available data and is by the factor $a/2L$ smaller than the one derived earlier \cite{fogler2005} for the case of 1D hopping. Thirdly, the characteristic fluctuation of currents for small samples, $\delta J\sim J_U\xi _m\sqrt{a/6L}$ predicted by Eq. (\ref{eq:avdis1}) exponentially increases with bias, consistent with the observations and opposite to the prediction (for 1D hopping \cite{fogler2005}) of bias suppressed variations.

\section{Discussion}
According to Sec. \ref{sec:small}, small samples should exhibit the Poole type of non-ohmicity, as opposed to the large samples that are expected to obey the Frenkel law. We would like to emphasize that the physics (of disordered systems) underlying the latter conclusions is dramatically different from the original hypotheses behind PF laws.

A number of numerical verifications have been provided in the literature for Poole law leading to the estimates of $a$ in the range of several nanometers, \cite{hill1971,ielmini2007} which can be reasonably interpreted as the characteristic linear scale of disorder (say, grain radius). For the case of Frenkel law, one can compare the original Frenkel coefficient $C_F$ with $C\sim C_S$ of the percolation theory, $C_F/C\sim \sqrt{6q^2/(a\varepsilon V_m)}$. Assuming $\varepsilon\sim 10$ and $a\sim 3$ nm yields $6q^2/(\varepsilon a)\sim 0.3$ eV, which is comparable to the characteristic disorder amplitudes, $V_m\sim 0.3-1$ eV. Hence, it is hard to decide in favor of Frenkel vs. percolation theory based just on the coefficient values.

We have extracted $C_F$ from a number of publications \cite{ismail2015,kim2014,hu2014,huang2014,schulman2015,slesazeck2015,song2018,lim2000,yuan2017} and found these coefficient varying by approximately one order of magnitude between different cases. Such variations are hardly attributable to the effect of $\varepsilon$, while they can be readily explained by the differences in the disorder parameters $V_m$ and $a$. Therefore we can cautiously favor the percolation theory vs. the original Frenkel argument.

Also, we note the prediction of log-normal probabilistic distribution of currents in small RRAM and PCM that has been observed (see \cite{karpov2017} and references therein). Furthermore, based on the above theory and following the existing approaches to mesoscopic systems \cite{raikh} one can describe the statistics of currents in small samples vs. bias, temperature, material disorder, and sample dimensions, to be presented elsewhere.

\section{A pathway to multi-valued memory}

A unique feature of the above described systems is that, in response to the applied bias, they  successively modify their constituting microscopic resistors. While this has been long realized for reversible modifications \cite{shklovskii1979}, a variety of new materials in modern technology of PCM and RRAM renders that phenomenon a new twist. The possibility of long-lived structural transformations makes such bias-induced modifications candidate elements of multi-valued nonvolatile memory.

\begin{figure}[t!]
\includegraphics[width=0.27\textwidth]{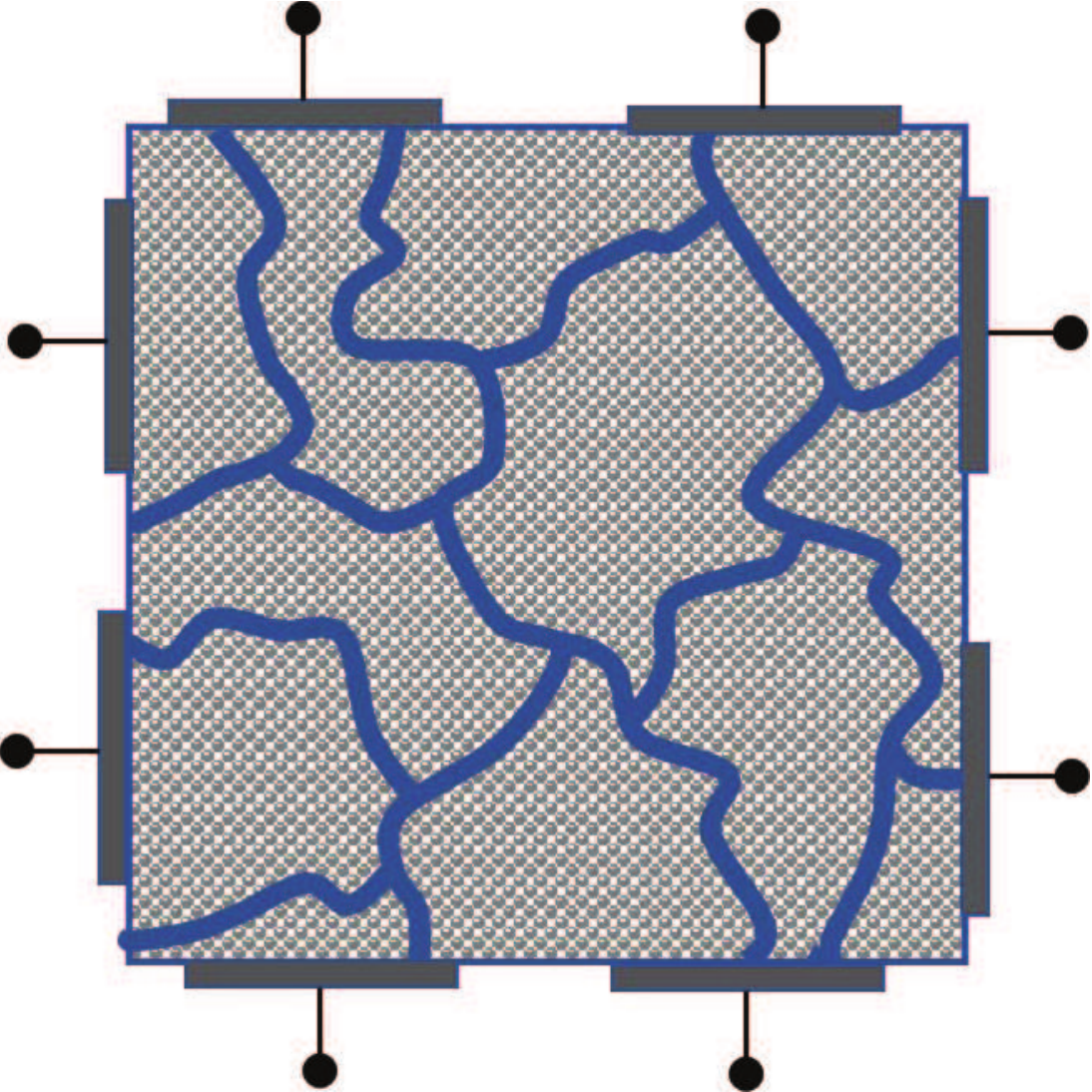}
\caption{A sketch of conductive pathways connecting multiple electrodes in a proposed design of multi-valued memory showing a fragment of the percolation cluster and multiple external electrodes. \label{Fig:PF_NVM}}
\end{figure}

While various architectures can be considered, here, we briefly illustrate the case of large devices having multiple ($M\gg 1$) electrodes sketched in Fig. \ref{Fig:PF_NVM}. Each electrodes can be connected to several or zero conductive pathways. If a certain bias is applied between any two electrodes, their connecting path will change its resistance to a random, but unique value, as described in Sec. \ref{sec:rev}.

The number of perceptive pathways can be estimated as $N=M!\approx \exp(M\ln M)\gg1$. For example, assuming $M=10$ yields an astronomical number of records, $N\sim 10^{10}$, if implemented with, say, $L\sim 1$ cm size sample and $\sim 1$ mm size electrodes. Hence, the memory capacity increases exponentially with the size of the model.

Conductive pathways connecting various pairs of the electrodes and sharing the same portion of infinite cluster will be mutually affected by a single bias-induced change. Furthermore, two subsequent pulses (${\cal E}_1$ and ${\cal E}_2$) applied to a given pair of electrodes produce a unique sequence of resistance changes resulting in an activation energy modification by  $\Delta V_{12} =[V_m{\cal E}_2qa(1-(V_m{\cal E}_1qa/V_m)^{1/2}]^{1/2}$. Therefore, two signals can be isolated sequentially in time. In summary, the system has a potential of recognition of binary spatiotemporal patterns.

The above model remains rather sketchy and lacking multiple details, such as specific algorithms of information processing and record erasing mechanisms (conceivably, by moderate heating). Yet, it introduces a concept of high capacity, distributed, single-trial learning model of storage, retrieval and recognition resembling some properties of the cortex of the mammalian brain.

\section{Conclusions}
We have shown that commonly observed Poole-Frenkel non-ohmic conduction in materials of non-volatile memory can be explained by the percolation theory corresponding to the cases of respectively small and large samples. In this framework, the applied bias modifies a limited number of microscopic resistances forming the percolation cluster. This understanding opens a pathway to multi-valued non-volatile memory and unique types of artificial neural networks with properties resembling that of natural ones.
\section*{Acknowledgements}
We are grateful to M. E. Raikh, B.I. Shklovskii, and A. V. Subashiev for useful discussions.


\begin{thebibliography}{99}
\bibitem{poole1916}H. H. Poole, Lond. Edinb. Dubl. Phil. Mag., 33, 112 (1916); Ibid., {\bf 34}, 195 (1917). Quoted in Ref. \onlinecite{hill1971}
\bibitem{hill1971}R. M. Hill, Poole Frenkel conduction in amorphous solids, Phil. Mag., {\bf 23}, 59 (1971)
\bibitem{frenkel1938}J. Frenkel, On Pre-Breakdown Phenomena in Insulators and Electronic Semi-Conductors, Phys. Rev. {\bf54}, 647-648, (1938).
\bibitem{aladashvili1989}D.I. Aladashvili, Z.A. Adamiya, K.G. Lavdovskii, E.I. Levin, and B.I.Shklovskii, Poole-Frenkel effect in the hopping conduction range of weakly compensated semiconductors, Fiz. Tekh. Poluprovodn, {\bf 23}, 213 (1989) [Sov. Phys. Semicond. {\bf 23}, 132 (1989)
\bibitem{perel}V. N. Abakumov, V. I. Perel, I. N. Yassievich, {\it Nonradiative Recombination in Semiconductors (Modern Problems in Condensed Matter Sciences)}, North-Holland (1991).
\bibitem{ielmini2007}D. Ielmini, and Y. Zhang, Evidence for trap-limited transport in the subthreshold conduction regime of chalcogenide glasses, Appl. Phys. Lett. {\bf90}, 192102, (2007).
\bibitem{nardone2012}M. Nardone, M. Simon, I. V. Karpov, and V. G. Karpov, Electrical conduction in chalcogenide glasses of phase change memory, J. Appl. Phys. {\bf 112}, 071101 (2012)
\bibitem{efros}B. I. Shklovskii, A. L. Efros, {\it Electronic Properties of Doped Semiconductors}, Springer, 1984.

\bibitem{shklovskii1979}B. I. Shklovskii, Percolation mechanism of electrical conduction in strong electric fields, Soviet Physics: Semiconductors, {\bf 13}, 53 (1979) [Fiz. Tekh. Poluprovodn. {\bf 13}, 93 (1979)]
\bibitem{levin1984}E. I. Levin, Percolation non-ohmic conductivity of polycrystaline semiconductors, Soviet Physics: Semiconductors, {\bf 18}, 158 (1984) [Fiz. Tekh. Poluprovodn. {\bf 18}, 255 (1984)]
\bibitem{shik}A. Y. Shik, {\it Electronic properties of inhomogenious semiconductors}, Gordon and Breach (1995).

\bibitem{fogler2005}M. M. Fogler and R. S. Kelley, Non-Ohmic Variable-Range Hopping Transport in One-Dimensional Conductors, Phys. Rev. Lett., {\bf 95}, 166604-8 (2005)
\bibitem{karpov2017}V.G. Karpov and D. Niraula, Log-Normal Statistics in Filamentary RRAM Devices and Related Systems, IEEE El. Dev. Letts., {\bf 38}, 1240 (2017).
\bibitem{ismail2015}M. Ismail, E. Ahmed, A. M. Rana, I. Talib, T. Khan, K. Iqbal and M. Y. Nadeem, Role of tantalum nitride as active top electrode in electroforming-free bipolar resistive switching behavior of cerium oxide-based memory cells, Thin Solid Films {\bf 583}, 95 (2015).
\bibitem{kim2014}W. Kim, S. Park, Z. Zhang and S. Wong, Current Conduction Mechanism of Nitrogen-Doped AlOx RRAM, IEEE Transactions on Electronic Devices {\bf61}, 2158 (2014).
\bibitem{hu2014}W. Hu,L. Zu, R. Chen, X. Chen, N. Qin, S. Li, G. yang and D. Bao, Resistive switching properties and physical mechanism of cobalt ferrite thin films, Appl. Phys. Lett. {\bf 104}, 143502 (2014).
\bibitem{huang2014}H.-P. Huang and S. Jou, Resistive Switching in TaN/AlNx/TiN Cell, International Journal of Chemical and Molecular Engineering {\bf8}, 607, (2014).
\bibitem{schulman2015}A. Schulman, L. F. Lanosa and C. Acha, Poole-Frenkel effect and variable-range hopping conduction in metal/YBCO resistive switching devices, J. Appl. Phys. {\bf118}, 044511 (2015).
\bibitem{slesazeck2015}S. Slesazeck, H. M\"{a}hne, H. Wylezich, A. Wachowiak, J. Radhakrishnan, A. Ascoli, R. Tetzlaffb and T. Mikolajickab, Physical model of threshold switching in NbO$_2$ based memristors, RSC Adv. {\bf5}, 02318, (2015).
\bibitem{song2018}S. Song, K. Kim, K. H. Jung, J. Sok and K. Park, Properties of Resistive Switching in TiO$_2$ Nanocluster-SiO$x(x<2)$ Matrix Structure, Journal of Semiconductor Technology and Science {\bf18}, 108, (2018).
\bibitem{lim2000}E. W. Lim and R. Ismail, Conduction Mechanism of Valence Change Resistive Switching Memory: A Survey, Electronics {\bf4}, 586, (2015).
\bibitem{yuan2017}F. Y. Yuan, N. Deng, C. C. Shih, Y. T. Tseng, T. C. Chang, K. C. Chang, M. H. Wang, W. C. Chen, H. X. Zheng, H. Wu, H. Quian and S. M. Sze, Conduction Mechanism and Improved Endurance in HfO2-Based RRAM with Nitridation Treatment, Nanoscale Res Lett.  {\bf12}, 574, (2017).
\bibitem{raikh}M. E. Raikh and I. M. Ruzin, {\it Transmittancy fluctuations in randomly non-uniform barriers and incoherent mesoscopic}in {\it Mesoscopic Phenomena in Solids}, edited
by B. L. Altshuller, P. A. Lee, and R. A. Webb, (Elsevier, New York,
1991), p. 315.




%
%
%
%
%
%
%
%
%
%


\end{thebibliography}
\end{document}